\documentclass[preprint,12pt]{elsarticle}

\usepackage{graphicx}
\usepackage{url}
\usepackage[breaklinks]{hyperref}

\usepackage{amssymb}
\usepackage{amsmath}

\journal{Chemical Engineering Journal}

\begin{document}

\begin{frontmatter}



\title{A temperature dependent framework to predict and control physical pellet quality in biomass extrusion}

\author[label1]{Richard T. Benders\corref{cor1}}
\ead{rtbenders@gmail.com}

\author[label2, label1]{Joshua A. Dijksman}

\author[label3]{Thomas M.M. Bastiaansen}

\author[label1]{Raoul Fix}

\author[label1]{Jasper van der Gucht}

\author[label4,label3]{Menno Thomas}

\affiliation[label1]{organization = {Physical Chemistry and Soft Matter Group, Wageningen University \& Research},
country={The Netherlands}}
\affiliation[label2]{organization = {Van der Waals-Zeeman Institute, Institute of Physics, University of Amsterdam},country={The Netherlands}}
\affiliation[label3]{organization = {Animal Nutrition Group,  Wageningen University \& Research},country={The Netherlands}}
\affiliation[label4]{organization = {Zetadec},addressline={Nudepark 73A},city={Wageningen},country={The Netherlands}}

\cortext[cor1]{Corresponding Author}




\begin{abstract}
The pellet manufacturing of biomass such as food and feed and bioenergy, various ingredients in powder or particle form are pressed together into a dense product, a pellet, with better nutritional, calorific, and handling properties than the individual particle-based input ingredients themselves. This makes pelleting valuable for up-converting industrial co-products from agriculture, forestry, food, and bio-energy sectors into higher-value products. However, processing particulate ingredient streams presents an industrial challenge and raises the important scientific question: ``Under which process conditions do loose particles bind together to form a mechanically rigid and durable pellet?"

This work provides new answers to this old research question in the context of biomass extrusion. We determine causal relationships between processing parameters and physical pellet quality through systematic pelleting experiments. Experimental results reveal that the interplay of steam conditioning temperature, production rate, and die geometry can be understood within an overarching framework of process interactions. 

Our framework introduces the concept of the ``stickiness temperature," $\mathrm{T^*}$, marking the onset of critical enthalpic reactions necessary for agglomeration of the individual particles within a pellet. Hence $T^*$ represents the boundary condition for inter-particle bond formation. We demonstrate how $T^*$ is achieved through a combination of steam conditioning and friction, and how these conditions can be controlled by adjusting process parameters.

Our findings emphasize the significance of pellet temperature in conjunction with die residence time for optimizing physical pellet quality while reducing specific energy consumption (J/kg). Validation through several trials and existing literature data confirms that our framework provides practical handles to intelligently enhance pelleting process efficiency and sustainability.

This work facilitates the transition towards a more circular economy by enabling production of economically viable and useful products from diverse ingredients by supplying additional operational parameters which may help in reducing energy consumption and the reduction of green house gases.
\end{abstract}







\end{frontmatter}



\section{Introduction}
\label{sec:intro}
Pellet manufacturing is an agglomeration technique widely employed across multiple industries to convert loose particulate materials into dense, consolidated products. In this process, diverse ingredients are first ground to sub-millimeter sized particles and thoroughly mixed. This mixture is then transformed through sequential processing steps into high-density, cylindrical pellets several millimeters in diameter.  These processing steps typically consist of steam treatment, consolidation, extrusion through specialized dies, and subsequent drying \cite{Thomas1997}.

The pelleting process involves complex energy transfer mechanisms that drive particle agglomeration. During steam conditioning, water vapor condenses on particle surfaces, releasing latent heat through enthalpy changes that raise mixture temperature and plasticize ingredients \cite{Thomas2020}. Subsequently, during consolidation and extrusion, friction between particles and between particles and the die dissipates additional thermal energy while mechanically compressing the material. These combined thermal and mechanical effects create conditions for inter-particle bond formation, transforming loose powder into cohesive pellets \cite{Thomas2020}.

However, the heterogeneous composition of ingredient mixtures introduces significant processing challenges. Different materials exhibit varying chemical structures that affect their thermodynamic properties, coefficient of friction, and response to steam conditioning. For instance, lignocellulosic materials respond differently to thermal treatment compared to protein-rich components, leading to mixture-specific processing behaviors \cite{Bastiaansen2023a,Bastiaansen2024,Bastiaansen2022}. These variations create complex interdependencies between ingredient properties and process parameters, making it difficult to predict optimal processing conditions.

The challenge is further compounded by confounding effects between ingredient characteristics and machine operation. For example, when comparing pellets made with low bulk density materials like wheat straw to conventional ingredients, results are often confounded. The low bulk density slows production rate due to reduced flow capacities inside the production line, increasing die residence time, which improves pellet quality \cite{Bastiaansen2023a}. This raises the question of whether quality improvement stems from the ingredient itself, the extended processing time, or both.

Furthermore, many process parameters depend on the configuration of the production plant and are influenced by equipment specifications such as motor characteristics and die geometry. These details are rarely reported, making it difficult to compare results between academic studies or across industrial facilities with different configurations. 

\subsection{The need for a coherent framework}
To overcome the engineering challenges in biomass pellet manufacturing, it is essential to develop a framework that is independent of both plant configuration and ingredient composition. Our framework provides an integrated view of the pellet manufacturing process, enhancing the understanding of process-specific parameters and enabling meaningful comparisons across different production plants. By adopting this approach, we can improve production efficiency in terms of both energy consumption and pellet quality while elucidating the fundamental physics governing bond formation during biomass extrusion.

To arrive at our framework, we systematically modified process conditions—such as steam conditioning temperature, production rate, and die geometry—during experimental trials. We used a ring-die pellet extruder to validate existing studies and uncover new insights into the physics of pellet manufacturing. These insights are captured within a novel process technological framework that introduces the concept of a ``stickiness temperature" ($T^*$), which is critical for binding loose ingredients into durable pellets.

Our systematic analysis addresses the fundamental questions: ``Under which process conditions do ingredients stick together to form a pellet?" and, from a physics perspective, ``What are the relevant boundary conditions for particle agglomeration?" To answer these questions, we also consider: ``When does agglomeration fail?"

We find that reaching $T^*$—the temperature at which particles begin to adhere—can be achieved through two mechanisms: 
\begin{itemize}
    \item Steam conditioning of the mixture prior to pelletizing.
    \item Frictional heating as the material is pressed through the pelleting die.
\end{itemize}
Our results show that pellet quality, measured by the pellet durability index (PDI), is primarily governed by the pellet's temperature achieved within the die.

Our proposed framework prioritizes plant-independent parameters such as the die residence time over production rate, and the net mechanical energy consumed during pelleting over the total mechanical energy consumed during pelleting. Furthermore, our framework offers a novel, data- and model-driven approach for optimizing the pelleting process, particularly when dealing with rapidly changing and complex ingredient compositions. By actively adjusting energy use in the pellet mill—balancing steam input and friction forces in the die according to ingredient properties—this method allows for more efficient and flexible production.

\begin{figure}[!th]
\centering
\includegraphics[width=0.65\textwidth]{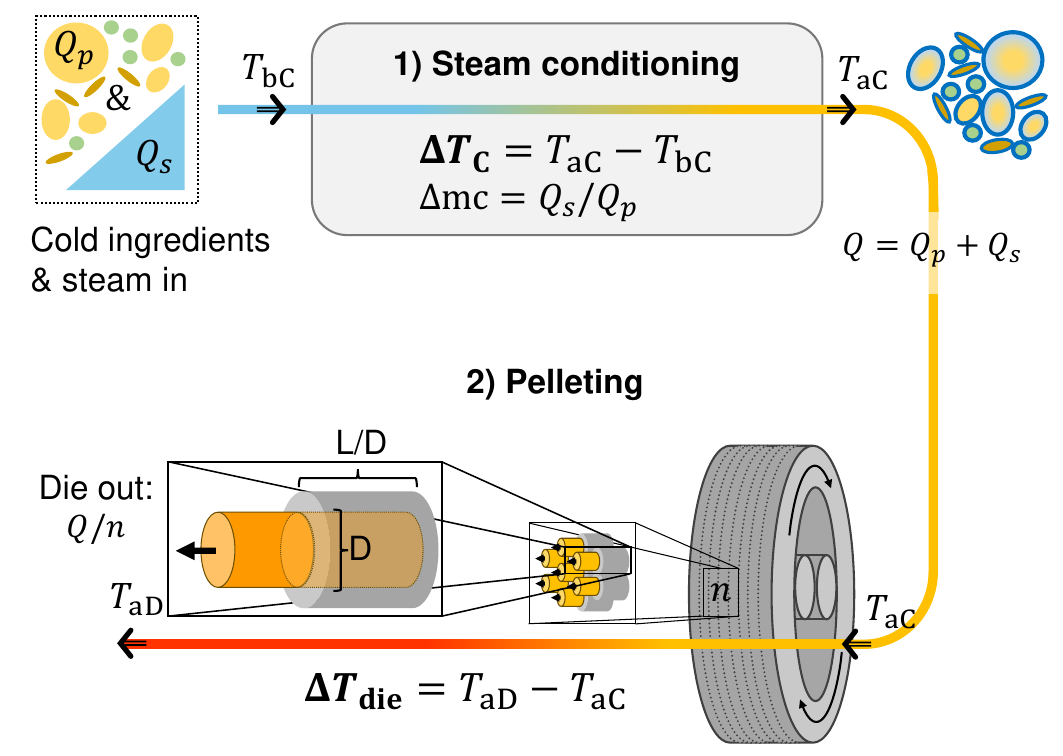}
\caption{\textbf{A schematic representation of the conditioning and pelleting process.} 
Cold ingredients enter the conditioner at $T_\mathrm{bC}$. The ratio of the steam injection rate, $Q_s$, to the ingredient flow rate, $Q_p$ (in $\mathrm{kg\ h^{-1}}$), determines the moisture and heat transfer during conditioning. During this process, latent heat ($H_\mathrm{vap}$) is transferred, increasing both the temperature and moisture content of the mixture (see Appendix \ref{app:sec:conditioning}). The conditioned material, now at temperature $T_\mathrm{aC}$, is compressed between the rollers and die before being extruded through one of the die holes. Friction within the die further increases the material's temperature from $T_\mathrm{aC}$ to $T_\mathrm{aD}$, the temperature measured immediately after extrusion. As the process operates in a closed system, the production rate ($Q$) is determined by the combined ingredient ($Q_p$) and steam flow rates ($Q_s$). The total number of holes in the pelleting die is specified as $n$, making the average flow rate per die hole $Q/n$. Finally, PDI is measured after the pellets are dried for 15 minutes. A 3D rendering of a pelleting die is shown in Fig. \ref{fig2:app}, and the schematic provided is not to scale. }\label{fig1:main}
\end{figure}

\section{The pelleting process and its parameters}
\label{sec:process}
Pellet production involves blending and processing ground particles from one or more ingredients to create durable pellets, e.g., with a typical goal of achieving a pellet durability index (PDI) of 95\% or higher while maintaining low production costs. The whole process can be divided into three key stages\cite{Thomas2020}:
\begin{enumerate}
    \item Steam conditioning
    \item Pelleting
    \item Cooling
\end{enumerate}  
During each of these stages key metrics are monitored, including gross and net energy use, and process temperatures. After cooling, the PDI, a measure of pellet robustness \cite{Thomas1996,Fahrenholz2012}, is determined to identify the relationships between processing settings and pellet quality. The PDI is reported as a fraction from 0 to 1, where a higher value signifies stronger, more durable pellets, which are resistant to abrasion during handling and transport, thus minimizing dust formation. We note that our framework relies on a specific type of pellet robustness test, where particle cohesiveness has a particularly strong influence on Holmen durability, in contrast to pellet hardness, which is typically assessed through various compression tests \cite{Thomas1996}. Holmen durability testing, which measures the amount of dust produced from pellets, remains one of the most widely used industrial metrics for monitoring physical pellet quality \cite{Thomas1996,Fahrenholz2012}. Fig. \ref{fig1:main} provides a schematic overview of the process and the registered process temperatures and moisture contents.

To establish our fundamental framework on pellet manufacturing, this study focuses on an ingredient mixture of maize and sugar beet pulp ($50/50\%$ w/w) as a reference material for compound animal feed production (unless specified otherwise). We confirm the generality of our framework by comparison with well-known literature data from Skoch \textit{et al}. \cite{Skoch1981}, the only work known to provide an almost complete report on their process conditions, data from Wecker \textit{et al.} \cite{Wecker2020} and an additional validation trial performed by us using a different ingredient composition.

Based on Fig \ref{fig1:main}, we identify three key industrial process variables that are most relevant to pellet production: 
\begin{itemize}
    \item Steam use during conditioning ($Q_s/Q_p$)
    \item Production rate ($Q$)
    \item Die geometry ($n$, $D$, $L$)
\end{itemize}
The latter two are plant- or configuration-dependent, but when combined, they form an plant-independent control parameter known as die residence time ($t_\mathrm{die}$). Die residence time is an intrinsic characteristic of the process that can be compared across different plants, unlike the isolated effects of production rate and die geometry, as further discussed in section \ref{sec:residence-time}. In Section \ref{sec:process}, we first describe the impact of these three tuning parameters on the physical pellet quality. In Section \ref{sec:tuning}, we then integrate these findings into a unified framework, defining the boundary conditions for pellet agglomeration. We emphasize the critical role of the  stickiness temperature $T^*$, identifying it as the onset temperature for successful pellet agglomeration.

\subsection{The effect of steam conditioning}\label{subsec:conditioning}
To establish the effectiveness of our experimental approach, we first demonstrate that we can replicate the results obtained by Skoch \textit{et al.} \cite{Skoch1981}, examining the effect of steam conditioning on PDI. During steam conditioning, hot steam condenses on the surface of relatively cold particles until thermal equilibrium is reached (see \ref{app:sec:conditioning}). The amount of condensed water can be gradually increased by carefully regulating the steam injection rate ($Q_\mathrm{s}$) relative to the particle mixture's flow rate ($Q_p$), supplied from the bunker by a feeder screw (Fig. \ref{fig1:main} and Eq. \ref{eq:app_1}). However, unlike Skoch's study, we conducted two experimental trials, using the same ingredients but with different mixture temperatures before steam conditioning (bC): once in summer ($T_{\mathrm{bC}}=24.3{}^\circ\mathrm{C}$) and once in winter ($T_{\mathrm{bC}}=11.8{}^\circ\mathrm{C}$).

\begin{figure}[!th]
\centering
\includegraphics[width=0.65\textwidth]{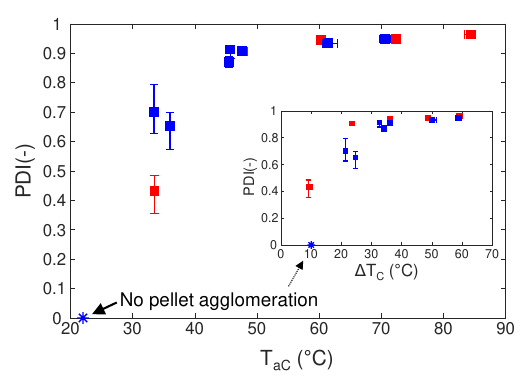}
\caption{\textbf{Increasing steam use during steam conditioning improves PDI}. Pellets were produced at a constant rate (approx. $250\ \mathrm{kg\ h^{-1}}$) and using the same die ($D=6\ \mathrm{mm}$ and $L/D=12$) from two maize-sugar beet pulp reference mixtures processed during summer (red) and winter (blue), differing in initial mixture temperature by $12.5{}^\circ\mathrm{C}$. Adjustment of $T_\mathrm{aC}$, induced varying $\Delta T_{\mathrm{C}}$ values (see inset), enabling a seasonal comparison between the effect of absolute temperature and relative temperature gain due to steam conditioning. In general, PDI increases with increasing steam use during conditioning. In winter, production of pellets with a $\Delta T_{\mathrm{C}}\approx10{}^\circ\mathrm{C}$, corresponding to a $T_{\mathrm{aC}}$ of $22{}^\circ\mathrm{C}$ was not possible (no agglomerates were formed) due to the lower initial temperature of the particle mixture, highlighted by the blue asterisks. Maintaining a $\Delta T_{\mathrm{C}}\approx10{}^\circ\mathrm{C}$ ($T_{\mathrm{aC}} \approx 34{}^\circ\mathrm{C}$), in summer however, does enable pellet agglomeration (see inset). Hence, there is an important difference between the effect of $T_\mathrm{aC}$ and $\Delta T_{\mathrm{C}}$ during the formation of a durable pellet.} \label{fig2:main}
\end{figure}
By performing the experiment twice, we can distinguish between the effects of the absolute temperature after conditioning ($T_\mathrm{aC}$) and the relative temperature increase during conditioning ($\Delta T_\mathrm{C}$) on PDI. While $T_\mathrm{aC}$ determines the absolute temperature of the particle mixture, $\Delta T_\mathrm{C}$ controls the amount of water that condenses during conditioning, which in turn influences the friction and heat generation within the die \cite{Skoch1981,Benders2024}. The combined results of both trials are presented in Fig. \ref{fig2:main}.

As expected, increasing steam use generally improves pellet PDI, as shown in Fig \ref{fig2:main}. These results align with those of Skoch \textit{et al.} \cite{Skoch1981} and common practices in pelleting factories. However, despite using the same die at a constant production rate, PDI is not solely influenced by the temperature reached during steam conditioning ($T_{\mathrm{aC}}$). There is a noticeable difference in PDI for pellets produced at $\Delta T_\mathrm{C}<35{}^\circ\mathrm{C}$ using mixtures with different initial temperatures $T_\mathrm{bC}$. This difference in PDI diminishes for $\Delta T_\mathrm{C}\geq35{}^\circ\mathrm{C}$. At constant $Q$—and therefore constant $t_\mathrm{die}$—the observed differences in PDI are attributed to variations in heat generation within the die ($\Delta T_\mathrm{die}$, see Fig. \ref{fig1:main}). Inside the die, the combined effects of steam and friction regulate the pellet temperature $T_\mathrm{aD}$, which ultimately determines physical pellet quality, as discussed in more detail below in section \ref{sec:tuning}.

\begin{figure*}[!th]
\centering
\includegraphics[width=1\textwidth]{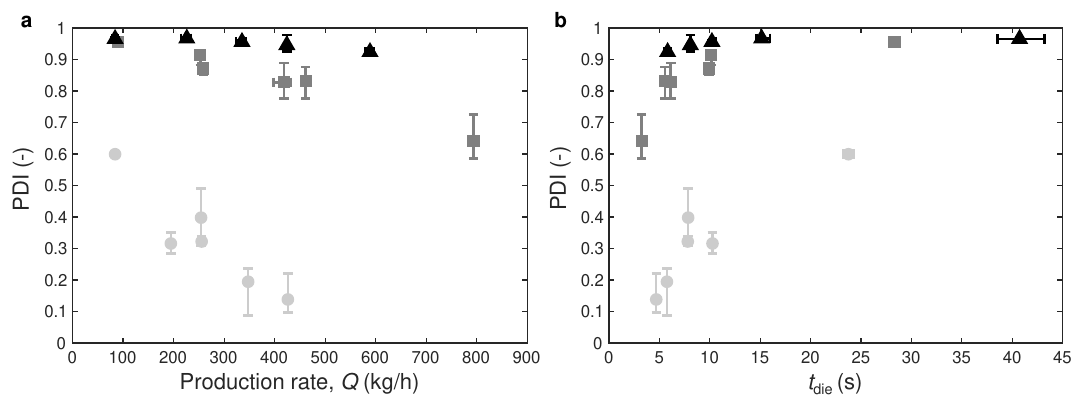}
\caption{\textbf{Die residence time, $t_{\mathrm{die}}$, can be used as tuning parameter in controlling physical pellet quality.} Pellets were produced using three different ring-dies (light gray sphere: $L/D=9.3$, gray square: $L/D=12$, black triangle: $L/D=16$) with $D=6\ \mathrm{mm}$, maintaining a constant $T_{\mathrm{bC}} (12.6\ \pm 0.2 {}^\circ\mathrm{C})$ and $\Delta T_{\mathrm{C}} (33.1\ \pm 0.5 {}^\circ\mathrm{C})$ at variable production rates ($Q$) (\textbf{a}) to alter the residence time within the die ($t_\mathrm{die}$) (\textbf{b}). PDI decreases with increasing $Q$, while PDI increases with longer residence time, due to the inverse relationship between $Q$ and $t_\mathrm{die}$. Understanding how $Q$ and $t_\mathrm{die}$ affect PDI across different ingredient compositions is key to optimizing pellet quality. A process operator can increase $Q$ up to the press engines' maximum power rating, maximizing throughput, minimizing specific mechanical energy usage, and reducing production costs. However, increasing $Q$ may decrease $t_\mathrm{die}$, likely reducing the mechanical quality of the pellets.}\label{fig3:main}
\end{figure*}

\subsection{The effect of production rate, die geometry and die residence time}\label{sec:residence-time}
In addition to steam conditioning, production rate, and die geometry are key parameters for producing durable pellets at low operational costs. Together, they determine die residence time—the duration ingredients remain in the die. While production rate and die geometry vary across production lines, residence time is plant-independent and enables cross-line comparisons. Notably, different combinations of production rates and die geometries can yield the same residence time. However, die geometry also affects the frictional surface area ($A_\mathrm{die}=n\pi D L$, for $n$ number of die channels with open ends) and the surface-to-volume ratio ($A_\mathrm{die}/V_\mathrm{die}=4/D$), influencing lubrication and heat generation (see section \ref{sec:tuning}). Dies with larger hole diameters, for example, are generally more energy-efficient \cite{Sultana2010}. A systematic experimental approach is needed to isolate the effects of these parameters.

During our pelleting trials, steam-conditioned particle mixtures are extruded through a ring-die containing cylindrical holes ($n$) with a specified diameter ($D$) and aspect ratio (also known as the compression ratio, $L/D$), where $L$ is the length of the extrusion channel (see \ref{fig1:main}). These geometric parameters, along with the production rate $Q$, dictate the average \mbox{residence} time ($t_{\mathrm{die}}$) of the pellet within the die (see methods \ref{MnM}). Since these parameters are interdependent, systematically varying the die geometry and production rate enables us to study the effect of residence time on PDI while managing their interrelationship. In our experiments, we limited ourselves to dies with $D=6\ \mathrm{mm}$ and $n=350$, maintaining a constant surface-to-volume ratio to avoid introducing confounding effects.

The parameter $t_{\mathrm{die}}$ is a critical optimization parameter. While longer die-residence time improves PDI \cite{Evans2019,Saensukjaroenphon2019}, it reduces production rate and increases gross specific mechanical energy usage—the total energy required per kilogram of pellets, including idle-load energy. Reducing the production rate to increase residence time raises energy consumption, elevating production costs and carbon emissions. Thus, any optimization must balance production rate, energy usage, and pellet quality.

Fig. \ref{fig3:main}a shows how increasing the production rate negatively impacts PDI, regardless of the die configuration. Fig. \ref{fig3:main}b shows that this reduction in PDI is due to the inverse relationship between $t_{\mathrm{die}}$ and $Q$, as $t_\mathrm{die}\propto1/Q$. Notably, multiple combinations of production rates and die geometries can result in the same $t_{\mathrm{die}}$ within the die. For instance, doubling $L$ while simultaneously doubling $Q$ maintains a constant $t_{\mathrm{die}}$. While seemingly obvious, this relationship is often overlooked in pellet production, where the focus tends to be on maximizing throughput rather than considering residence times. Contrastingly, for a fixed die configuration, variations in $Q$ lead to changes in $t_{\mathrm{die}}$, which in turn affects PDI. This complicates data analysis from typical pelleting experiments, where the production rate is often determined by a fixed power consumption of the pellet mill's engines, as seen in ref. \cite{Wood1987}. Interestingly, the results in Fig. \ref{fig3:main}b suggest that PDI cannot be predicted solely by $t_\mathrm{die}$ and $T_\mathrm{aC}$. The observed differences in PDI are explained by the varying amounts of heat generated from die friction across different channel lengths ($L$), which influences the pellet temperature. This effect is discussed further in section \ref{sec:tuning}, where we describe how steam use during conditioning and die geometry can be modified to improve physical pellet quality through modulation of the process temperature.

\section{Temperature matters: setting the boundary for pellet agglomeration}\label{sec:temperaturematters}
\begin{figure}[!th]
\centering
\includegraphics[width=0.65\textwidth]{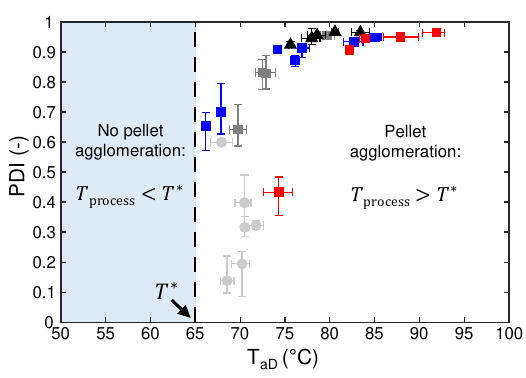}
\caption{\textbf{Pellet agglomeration does not occur below $T^*$, the boundary condition for stickiness.} The effects of all process parameters shown in Figures \ref{fig2:main}, \ref{fig3:main}a, and  \ref{fig3:main}b can be unified onto a single master curve by plotting PDI against pellet temperature immediately after extrusion. This suggests that pellet temperature is the primary determinant of PDI. The master curve indicates a minimum temperature threshold ($T^*$) for the maize-sugar beet pulp mixture, estimated at $65\ {}^\circ\mathrm{C}$. Below this temperature, agglomeration does not occur, as shown by the blue region. Above the \textit{stickiness temperature}, ingredients can agglomerate and form pellets. The color and symbols, representing the different process conditions, are matched to the data presented in the previous figures, furthermore the data presented by the red squares was reproduced from Benders \textit{et al.} \cite{Benders2024}.} \label{fig5:main}
\end{figure}

So far, we have described the effects of three key processing factors in pellet manufacturing— steam conditioning, production rate, and die geometry— and their relation to changes in pellet PDI. Unlike earlier work, we applied a consistent model ingredient composition across all trials, and demonstrated that each factor influences PDI. Crucially, our results show that the effects of initial mash temperature, steam conditioning, residence time, and die geometry all converge on a critical single parameter: the final pellet temperature ($T_\mathrm{aD}$), measured directly after the die, as shown in Fig. \ref{fig5:main}. Previously, Agar \textit{et al.}  \cite{Agar2018} reported a similar strong correlation between pellet temperature and physical pellet quality during the processing of various agricultural and forestry biomass ingredients. However,  This convergence underscores the importance of achieving a process temperature within the die that meets or exceeds the stickiness temperature ($T^*$), which governs whether agglomeration and successful pellet formation can occur. 

\subsection{Ingredients agglomerate when process temperature exceed the stickiness temperature}
From the results in Fig \ref{fig5:main}, we introduce the concept of a critical temperature threshold, referred to as the minimum agglomeration temperature or the \textit{stickiness temperature} ($T^*$), similar to the concept of stickiness in food science \cite{Palzer2005}. The stickiness temperature, estimated to be approximately $65\ {}^\circ\mathrm{C}$ for our model composition, marks where pellet agglomeration begins during extrusion. Below $T^*$, agglomeration does not occur, hence $T^*$ sets the physical and chemical boundary for successful pellet formation. 

\begin{figure}[!th]
\centering
\includegraphics[width=0.6\textwidth]{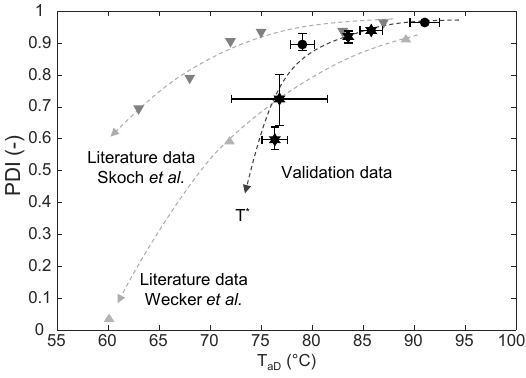}
\caption{\textbf{$T^*$ presents a generalized, composition-dependent agglomeration temperature in pellet manufacturing .} In general, PDI increases with increasing pellet temperature, similar to the results presented in Fig. \ref{fig5:main}. This is demonstrated using literature data from Skoch \textit{et al.} \cite{Skoch1981}, data from Wecker \textit{et al.} \cite{Wecker2020}, as well as in an additional validation trial. In the validation trial, a mixture of maize, soybean meal, and oat hulls was pelletized using a ring die with $6\ \mathrm{mm}$ diameter holes and compression ratios of $6$ (stars) and $9.3$ (spheres). During this trial, $\Delta T_\mathrm{C}$ was increased from approximately $45\ {}^\circ\mathrm{C}$ to $75\ {}^\circ\mathrm{C}$ to produce pellets at different process temperatures which are measured directly after the die ($T_\mathrm{aD})$. The results demonstrate the generalized concept of an ingredient-dependent $T^*$, which defines the process temperature at which ingredients begin to stick together and form pellets. $T^*$ is determined by the asymptotic relationship between PDI and $T_\mathrm{aD}$ indicated by the dashed arrows.}\label{fig6:main}
\end{figure}

Differences in ingredient composition or process conditions can shift the $T^*$ as the onset temperature. For example, the use of crops grown in different seasons or on different geographic locations, as well as the amount of water added during mixing or conditioning, a plasticizer for starch- or protein-containing ingredients, may increase or decrease $T^*$, indicated by the difference in curvature between the experiments run in summer (red) or winter (blue) in Fig. \ref{fig5:main}. Despite these differences, the concept of $T^*$ as the boundary condition for successful pellet agglomeration is not unique to a binary mixture of maize and sugar beet pulp. We confirm the generality of $T^*$ by comparing our findings with well-known literature data from Skoch \textit{et al}. \cite{Skoch1981}, data from Wecker \textit{et al.} \cite{Wecker2020}, and by comparison of our results with an additional validation trial performed by us using a different ingredient composition (Fig. \ref{fig6:main}). 

Based on figures \ref{fig5:main} and \ref{fig6:main}, we conclude that each ingredient mixture has its own unique $T^*$. As a result, producing durable pellets at the lowest operational cost requires specific tuning of the process conditions for each ingredient mix. Achieving $T^*$ can be done in two ways: through steam conditioning or friction. In a pelleting process where friction from the ring die is the primary source of heat generation, the following condition must be met (Eq. \ref{eq:main_1}): 
\begin{equation}\label{eq:main_1}
T_\mathrm{bC}+\Delta T_\mathrm{C}+\Delta T_\mathrm{die}\geq T^*
\end{equation}
While $\Delta T_\mathrm{C}$ is controlled by adjusting the process temperature in the conditioner ($Q_s/Q_p$, see section \ref{app:sec:conditioning}), $\Delta T_\mathrm{die}$ depends on the frictional work within the ring die. Tuning $\Delta T_\mathrm{die}$ requires a detailed understanding of the complex friction dynamics in the die. Additionally, thermal losses during the gravitational transport of material from the conditioner to the pelletizer may reduce the mixture’s temperature, thereby increasing the energy demand within the press to satisfy Eq. \ref{eq:main_1}. However, these losses are considered minimal in our pilot-scale pelletizer, where the distance between the conditioner and the pellet die is approximately $0.5\ \mathrm{m}$. In the next section, we explore the processing and ingredient factors that affect $\Delta T_\mathrm{die}$.

\subsection{Process conditions regulate die friction}\label{sec:tuning}
The pellet extrusion process is a complex, semi-continuous consolidation and extrusion operation. Within a pellet press, the rollers exert force onto the particle mixture, enabling ingredient flow through the die. The press engines drive the ring die's rotation at a constant speed, while the rollers inside remain fixed in position but rotate as the die and particle mixture rotate. This setup applies a compressive force from the rollers onto the powder mixture, pushing it into individual die-holes with each rotation of the die. As a result, extrusion from the producer’s perspective is continuous, but it appears intermittent (stop-and-go) within each die hole. The force exerted by the rollers in the normal direction of the die-holes ($F_n$) must overcome internal resistances, or friction forces ($F_\mu$), within the die and between particles to enable material flow and compaction into pellets. Consequently, the power provided by the press engines ($P$ in $\mathrm{W}$) scales proportionally with the extrusion force ($F_n$ in $\mathrm{N}$) and the material's sliding velocity through the die ($v$ in $\mathrm{m\ s^{-1}}$), which is directly related to the production rate, $Q$. Inside the die-holes, friction converts mechanical energy into heat, increasing the temperature and promoting particle bonding.

A detailed description of how extrusion forces and pressures develop during pellet extrusion is available elsewhere (see refs \cite{Holm2006,Holm2011}). However, the extrusion pressure, the pressure exerted by the rollers onto the ingredient mixture, depends on factors such as the die’s compression ratio and the coefficient of friction between the particle mixture and the die wall surface \cite{Holm2006,Holm2011}. This work focuses on two tunable aspects that increase the force and pressure required during extrusion, thereby affecting heat generation within the die: the coefficient of friction $\mu_\mathrm{app}$ and the die-hole surface area $A_\mathrm{die}$. The latter depends directly on the die's geometric specifications ($A_\mathrm{die}=n \pi D L$). When $D$ remains constant, $A_\mathrm{die}$ scales linearly with the length ($L$) of the channels. Hence, dies with a longer channel length $L$ increase the work of friction during pellet extrusion, reflected by the increased power consumption by the presses engines \cite{Benders2024}.

The coefficient of friction ($\mu_\mathrm{app}$) is influenced by several factors: the type of metal used to manufacture the die \cite{Murase1984}, the ingredient composition \cite{McKenzie1968}, and process parameters like the production rate \cite{McKenzie1968}, the run-time and wear of a particular die \cite{Sultana2010}, and the amount of steam provided \cite{Benders2024}. In earlier work, Benders et al. \cite{Benders2024} demonstrated how the key processing parameter—steam use during conditioning—affects the formation of lubrication layers during pellet extrusion. The thickness of these lubrication layers, relative to the surface roughness of the metal die, ultimately determines the coefficient of friction \cite{Benders2024,Wasche2014,Taylor2022,Popov2010}. Increasing steam use from $0.035$ to $0.053$ kg per kg of ingredients reduced the coefficient of friction by approximately $16\%$ when using dies with a diameter of $6\ \mathrm{mm}$ \cite{Benders2024}.

This steam-induced lubrication mechanism underscores our preference for using $A_\mathrm{die}$ over the $L/D$ ratio introduced by Holm \textit{et al.} \cite{Holm2006}, as $L/D$ does not account for surface-to-volume effects. While steam conditioning increases moisture content per unit volume, changes in $D$ alter the surface-to-volume ratio by $4/D$. A smaller $D$ reduces the amount of lubricant per unit area by the same factor, potentially thinning the lubrication layers and increasing energy consumption \cite{Sultana2010}. Therefore, our study focused exclusively on evaluating the effect of $L$. Additionally, a constant production rate $Q$ was maintained to provide a constant sliding velocity $v$, recognizing that the coefficient of friction $\mu_\mathrm{app}$ is influenced by this velocity \cite{McKenzie1968,Benders2024,Wasche2014}.

\begin{figure*}[!th]
\centering
\includegraphics[width=1\textwidth]{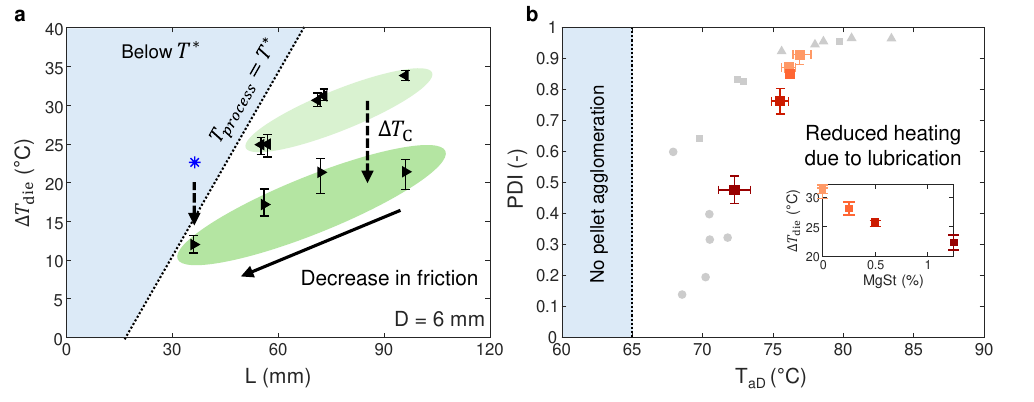}
\caption{\textbf{Various process conditions affect the heat generation within the the die $\Delta T_\mathrm{die}$ through changes in the work of friction, ultimately impacting physical pellet quality.} \textbf{a.} Pellets were produced at two different $\Delta T_\mathrm{C}$ levels, at a constant $Q$ (approx. $250\ \mathrm{kg\ h^{-1}}$), and with four different dies varying in their channel lengths ($L$). Lengthening the die channel increases the surface area over which friction acts, resulting in increased heating of the pellets during extrusion,  as reflected by $\Delta T_\mathrm{die}$. Additionally, increasing $\Delta T_\mathrm{C}$ from $31.8\pm0.6{}^\circ\mathrm{C}$ (left-pointing triangle) to $49.8\pm 0.8{}^\circ\mathrm{C}$ (right-pointing triangle) provides more steam-induced lubrication and consequently dissipates less heat during pellet extrusion, reducing $\Delta T_\mathrm{die}$. The blue asterisk marks a data point which does not meet the criteria of equation  \ref{eq:main_1}; therefore, pellet agglomeration was not possible under these process conditions. \textbf{b.} The effect of die friction on physical pellet quality is validated by adding magnesium stearate (MgSt) as a lubricant to the mixture at a constant production rate (approx. $250\ \mathrm{kg\ h^{-1}}$), $\Delta T_\mathrm{C}$ $30.7\pm0.5{}^\circ\mathrm{C}$ and $T_\mathrm{aC}$ $47.8\pm0.4{}^\circ\mathrm{C}$. MgSt reduces friction within the ring die and consequently impacts heat generation (inset). Increasing MgSt concentrations reduce $\Delta T_{\mathrm{die}}$, lowering $T_\mathrm{aD}$, which immediately reflects the reduction in PDI after production. The gray data are included for reference and correspond to the average PDI values of the pellets produced according to Fig. \ref{fig3:main}, using the same steam conditioning parameters: $\Delta T_{\mathrm{C}}\ 33.1\ \pm 0.5 {}^\circ\mathrm{C}$ and $T_\mathrm{aC}$ $45.7\pm0.5{}^\circ\mathrm{C}$.} \label{fig7:main}
\end{figure*}

In Fig. \ref{fig7:main}a we summarize how steam use, $\Delta T_\mathrm{C}$, and die length $L$, under a constant production rate $Q$ and die-hole diameter $D$, affect friction and subsequent heat generation within the die. The increase in temperature across the die ($\Delta T_\mathrm{die}$) captures the generated heat, and we demonstrate how friction plays a critical role in controlling process temperatures to satisfy Eq. \ref{eq:main_1}. Lastly, we apply this enhanced understanding of friction’s role in process temperature control in a final validation trial. In this trial (Fig. \ref{fig7:main}b), $\mu_\mathrm{app}$ was reduced through the addition of a model-type lubricant into the particle mixture, resulting in a decrease in die heat generation ($\Delta T_\mathrm{die}$) and a simultaneous reduction in pellet PDI.

The results in Fig. \ref{fig7:main}a show that increasing the die channel length ($L$) while keeping $n$, $D$, $Q$, and $\Delta T_\mathrm{C}$ constant leads to an increase in $\Delta T_{\mathrm{die}}$, along with an increase in the press’s energy consumption (see Fig. A.7. in ref. \cite{Benders2024}). This simultaneous rise in process temperatures and mechanical energy consumption results from increased frictional work within the die due to the larger surface area of the die holes ($A_\mathrm{die}$) as $L$ grows. This effect is demonstrated for two levels of $\Delta T_\mathrm{C}$, highlighted by the green clusters, which represent friction coefficients differing by approximately $16\%$ (see ref. \cite{Benders2024}).

Additionally, in Fig. \ref{fig7:main}a, we show that increasing steam use (while keeping $n$, $L$, $D$, and $Q$ constant) results in a reduction of $\Delta T_{\mathrm{die}}$, consistent with the findings of Skoch \textit{et al.}\cite{Skoch1981}. This reduction is explained by the formation of lubrication layers within the ring die, which decreases the coefficient of friction ($\mu_\mathrm{app}$) \cite{Benders2024}. As steam use during conditioning increases—illustrated by the light and dark green areas representing $\Delta T_\mathrm{C}$ values of $31.8{}^\circ\mathrm{C}$ and $49.8{}^\circ\mathrm{C}$, respectively—the coefficient of friction decreases, reducing heat generation and, consequently, lowering $\Delta T_\mathrm{die}$.

Together, the sum of $T_\mathrm{aC}$ and $\Delta T_\mathrm{die}$ determines whether pellet agglomeration will occur, based on the criteria in equation \ref{eq:main_1}. For example, our attempts to produce pellets from a particle mixture at $T_{\mathrm{post-conditioner}}\approx45{}^\circ\mathrm{C}$ using a ring die with $D=6\ \mathrm{mm}$ and $L/D=6$ failed (marked by the blue asterisk in Fig. \ref{fig7:main}a). We believe that the heat generated through friction within this die was insufficient to reach the required agglomeration temperature $T^*$. However, increasing $T_{\mathrm{aC}}$ allowed us to meet the requirements of equation \ref{eq:main_1} and initiate pellet agglomeration. At $T_{\mathrm{aC}}=61.4\pm 0.3 {}^\circ\mathrm{C}$, pellets were produced with a PDI of $0.09\pm 0.02$ (as shown in Fig. \ref{fig7:main}a), and further increasing $T_{\mathrm{aC}}$ to $72.8\pm 1.3{}^\circ\mathrm{C}$ raised the PDI to $0.44\pm 0.10$ (data not shown).

\subsection{Lubricant addition reduces die friction and lowers pellet quality}
Finally, to conclude our systematic analysis, we apply our findings in a final validation trial, demonstrating that PDI can be adjusted by changing the coefficient of friction ($\mu_\mathrm{app}$) of the ingredient mixture (Fig. \ref{fig7:main}b). Lubricants such as oils and fats, commonly added during mixing to ensure the nutritional composition of animal feed, also significantly impact pellet PDI. While steam conditioning improves PDI and reduces mechanical energy usage, oil- or fat-based lubricants tend to reduce PDI\cite{Thomas1998}—an effect not fully understood. Unlike steam, which increases both temperature and moisture content, oil or fat additives do not. The effect of lubricants on heat generation within the die and its relation to PDI was assessed by adding magnesium stearate (MgSt) to the particle mixture at concentrations up to  $1.25\%$ w/w. Magnesium stearate, a water-insoluble lubricant commonly used in pharmaceutical tabletting \cite{Perrault2010,Morin2013}, reduces $\mu_\mathrm{app}$ and subsequently decreases net mechanical energy use—representing energy dissipated during pellet extrusion—by $28.8\%$, from $53.0\ \mathrm{kJ\ kg^{-1}}$ ($14.7\ \mathrm{kWh\ t^{-1}}$) to $37.7\ \mathrm{kJ\ kg^{-1}}$ ($10.5\ \mathrm{kWh\ t^{-1}}$) as MgSt content increased from $0\%$ to $1.25\%$ w/w, with all other process parameters held constant. Concurrently, $\Delta T_\mathrm{die}$ decreased by approximately $8.5 {}^\circ\mathrm{C}$ due to reduced friction  (Fig. \ref{fig7:main}b, inset), resulting in lower physical pellet quality, as the die temperature ($T_\mathrm{aD}$) also decreased (Fig. \ref{fig7:main}b).

\section{Summary, outlook and considerations}\label{sec:summary}
The main objective of this study was to answer the question: ``Under which process conditions do particulate ingredients bind together to form a mechanically rigid and durable pellet?" Our systematic investigation shows how key common processing parameters, such as steam conditioning, production rate, die geometry, and the addition of lubricants, affect physical pellet quality in pellet manufacturing. The primary novel finding is the identification of the ingredient-dependent stickiness temperature, $T^*$, as a critical temperature threshold, below which agglomeration fails. Consequently, $T^*$ marks the temperature at which ingredients begin to agglomerate into pellets, establishing it as an essential parameter for pelleting process optimization. This observation was further supported by literature data and a validation trial, demonstrating the generality of our conceptual framework: ingredients start to stick together once their specific $T^*$ is reached within the pelleting process, due to supplied heat in the form of steam and friction.

Building on these insights, we described a framework that enabled a data-driven approach to optimizing the pelleting process, particularly when dealing with complex and rapidly changing ingredient compositions. One of the most challenging aspects is controlling $\Delta T_\mathrm{die}$, which is influenced by the manifold ways in which the apparent coefficient of friction ($\mu_\mathrm{app}$) can be affected by various ingredients and process conditions. For instance, different ingredient mixtures have distinct physical and chemical properties. One such property, which was not specifically evaluated in this work, is the specific heat capacity ($c_p$) \cite{Kulig2007}. The $c_p$ determines how much energy is required to heat the ingredients, consequently affecting how much steam transfers into the particle mixture during conditioning and how much frictional heat is needed to raise their temperature by a given $\Delta T$ during extrusion. Additionally, small quantities of fat or oil-rich components can reduce $\mu_\mathrm{app}$ \cite{Bastiaansen2025}, similar to the effect of magnesium stearate addition. Consequently, such a reduction in friction needs to be counterbalanced by adjustments in process conditions to reach process temperatures above $T^*$. Future work in upcoming papers by our group will focus on these aspects. 

Our work highlights the critical role of process temperature and opens the door to rethinking the design of pelleting equipment. Currently, pelleting dies rely solely on frictional heating to raise material temperatures after steam conditioning, but our findings suggest that temperature-regulated dies could offer new ways to fine-tune PDI. For instance, lubricating components like fats or oils reduce frictional heat generation, and during production, dies with a large $L/D$ are required to raise process temperatures above $T^*$ under conventional pellet mill designs. A temperature-controlled die, while presenting an engineering challenge, could serve as as an additional heating source to optimize physical pellet quality in such scenarios. However, under regular process conditions where frictional heating is sufficient, a temperature-regulated die may not significantly improve physical pellet quality \cite{Segerstrom2014}. Nevertheless, our framework provides a scientific basis for such innovations, potentially driving the next generation of pellet press designs capable of processing a wider variety of co-products and unconventional ingredient streams, thereby supporting the transition toward a more sustainable circular agricultural model. 

Our results also provide insight into some of the challenges encountered daily in industrial pellet manufacturing, such as the effects of seasonal changes on mill performance and product quality. Seasonal variations often alter the temperature of the initial ingredient mixture, requiring adjustments to both $\Delta T_\mathrm{C}$ and $\Delta T_\mathrm{die}$ to ensure that $T^*$ is reached. However, increasing $\Delta T_\mathrm{C}$ through higher steam usage can reduce the die's ability to heat the material, leading to lower $\Delta T_\mathrm{die}$ due to steam-induced lubrication.

Another notable challenge is identifying the individual causal parameters in the pelleting process, particularly within the pellet press itself. Our experiments show that when pellet temperatures approach $T^*$ (e.g., $T_\mathrm{aC} \approx 45.7 {}^\circ\mathrm{C}$ with a die diameter of $6\ \mathrm{mm}$ and a compression ratio of $\mathrm{L/D}=9.3$), die residence time significantly impacts PDI (light gray spheres in Figs. \ref{fig3:main}b and \ref{fig5:main}). However, when pellet temperatures exceed $T^*$, the influence of residence time diminishes, as shown by the black triangle data points in the same figures. This finding indicates that the impact of process conditions on pellet quality is contingent on other variables, leading to varied outcomes across different experiments. Nevertheless, the results itself align with those from food science, where bond formation rates accelerate with increasing temperature and force \cite{Palzer2005,Roos2015}. When process temperatures exceed $T^*$, bond formation occurs rapidly, and the residence time inside the die becomes less critical for pellet quality. 

While this study provides a robust framework for understanding and optimizing the pelleting process, certain factors—such as pressure during extrusion, particle size, and the post-production cooling and drying process—remain underexplored. These factors can influence pellet quality over time, particularly through moisture exchange with the environment. Probing the role of these parameters are fruitful avenues for future work in order to understand how ingredient-specific properties and time-dependent aging processes affect pellet durability, both immediately after production and during storage.

\section{Conclusion}
Our work establishes the critical role of process temperature in pellet production, particularly through the identification of $T^*$. By addressing the interdependencies between key process parameters, we have laid the foundation for a more efficient and adaptable approach to pellet manufacturing. This framework can guide future efforts to optimize both energy use and pellet quality, particularly as ingredient compositions continue to evolve.

\section*{Acknowledgments}
We wish to acknowledge Zorheh Farmani, Jose Ruiz Franco, Martijn van Galen, Ruud Holtermans, Reinier Kaanen, Sophie van Lange, Jordi Rijpert, Prince Njarnob Samoah and Jorik Schaap for assisting during the experimental investigation. We wish to thank Guido Bosch, Wouter Hendriks and Sonja de Vries for their contributions to the discussions, while developing the of the pelleting framework. This study was financially supported by The VICTAM Foundation, Aeres Training Centre International, Agrifirm NWE B.V., DSM, Elanco Animal Health, Feed Design Lab, Pelleting Technology Netherlands, Phileo S.I. Lesaffre, Topsector Agri \& Food, Zetadec, and Wageningen University \& Research, as partners in the project ``Pelleting in Circular Agriculture'' (project number: LWV1965).

\section*{Declarations}

\begin{itemize}
\item Availability of data and materials: all processed data is available in the main article and supplementary materials; raw data will be provided upon reasonable request.
\item Authors' contributions: 
\end{itemize}

\textbf{Richard Benders:} Conceptualization, Methodology, Investigation, Formal analysis, Data Curation, Visualization, Writing - original draft
\textbf{Joshua Dijksman:} Conceptualization, Methodology, Writing - review \& editing, Supervision.
\textbf{Thomas Bastiaansen:} Investigation, Writing - review \& editing.
\textbf{Raoul Fix:} Conceptualization, Investigation, Visualization
\textbf{Jasper van der Gucht:} Conceptualization, Writing - review \& editing, Supervision
\textbf{Menno Thomas:} Conceptualization, Methodology, Writing - review \& editing, Supervision, Project administration.

\appendix
\clearpage
\section{Methods}\label{MnM}
\subsection{Materials}
In total $18000\ \mathrm{kg}$ kg of the milled reference material (cleaned corn kernels and low-sugar sugar beet pulp $50/50\%$ w/w) was purchased from Research Diet Services (Wijk bij Duurstede, The Netherlands) in two separate orders of $7000\ \mathrm{kg}$ and $11000\ \mathrm{kg}$ respectively, as specified in ref. \cite{Benders2024}. The exact chemical compositions is unknown and may have differed between the first and second order, however, these variations were not the focus of this study, for which the chemical composition was not analyzed. The initial temperature of the ingredients was different during the summer ($T_{\mathrm{bC}}=24.3{}^\circ\mathrm{C}$) and during winter ($T_{\mathrm{bC}}=11.8{}^\circ\mathrm{C}$) trials, due to the difference in ambient temperature within the processing hall.

The validation mixture, of which the data is presented in Fig. \ref{fig6:main}, was also purchased from Research Diet Services, and contained maize, soy bean meal (high-protein) and oat hulls ($17$, $33$ and $50\%$ w/w respectively), ground over a $4$-mm screen using a hammer mill.

Magnesium stearate ($5\ \mathrm{kg}$ plastic drum, $3.8$-$5.0\%$ Mg) was ordered from Fischer Scientific and the specified quantities ($0.25\%$, $0.5\%$ and $1.25\%$ w/w) were mixed into the reference mixture using a Nauta Mixer (Hosakawa Micron B.V., Zelhem Netherlands) for $15$-$20$ minutes.

\subsection{Pellet production and monitoring process}
Pellets were produced using a RMP200 ring-die pelletizer (Münch Edelstahl GmbH, Hilden Germany), with a variable production rate between $80$ and $800\ \mathrm{kg\ h^{-1}}$ regulated by a feeder screw. Steam conditioning was performed as described in ref. \cite{Benders2024}, using high-quality super-heated steam, in order to increase the cold ingredient mixtures temperature ($T_\mathrm{bC}$), in order to to maintain a certain temperature after conditioning $T_\mathrm{aC}$ or a certain $\Delta T_\mathrm{C}$. The steam-flux was recorded using a H250 flow meter (Krohne, Duisburg, Germany) .

The RMP200 pelletizer was configured with two rollers, and a series of the ring-dies with $6$ $\mathrm{mm}$ diameter die-holes and a unique compression ratio ($L/D$; $6$, $9.33$, $12$ and $16$) as specified in ref \cite{Benders2024}. Each ring-die has $350$ holes, from which we can compute the volumetric flow rate and determine the ring-die residence time $t_\mathrm{die}$ (see below).

Electrical energy consumption was monitored according to the specifications in ref. \cite{Benders2024}. Each experimental run was initialized with a $20$ minutes period, during which feeder speed and steam injection rate were adjusted to obtain steady-state. During each run, approximately $2\ \mathrm{kg}$ of pellets were sampled directly after leaving the press. These samples were subsequently cooled at ambient air for $10\ \mathrm{min}$, transferred into plastic bags at room temperature and stored at $4$ ${}^\circ \mathrm{C}$ until pellet durability was analyzed (minimum storage time was $24$ hours after production, maximum storage time was $1$ week). The sampling procedure was replicated $2$ to $3$ times per run, with $15$-$20$ minute intervals.

Registered process parameters include ingredient mixture temperature and moisture content before and after steam conditioning, pellet temperature and moisture content after pellet extrusion, electrical current, and production rate ($\mathrm{kg\ h^{-1}}$). The temperature measurements were performed as described in \cite{Bastiaansen2023}, the moisture content during the summer trial was determined as described in \cite{Vego2023}, the moisture content during the winter trial was determined from the steam-flux data, and the production rate was determined as described in  \cite{Bastiaansen2023}, using a $2$ minute measurement period. From these recordings the change in moisture content ($\Delta \mathrm{mc}$, in kg water per kg ingredients) during the steam conditioning process, the change in temperature during conditioning ($\Delta T_\mathrm{C}$) and the change in temperature over the ring-die ($\Delta{T_\mathrm{die}}$) were calculated. 

The error-bars in the figures represent the minimum and maximum values e.g., for $T_\mathrm{aD}$ or PDI, or the minimum and maximum changes in the process conditions e.g., the minimum $\Delta T_\mathrm{die}$ is obtained by subtracting the maximum $T_\mathrm{aC}$ (the highest temperature after conditioning) from the minimum $T_\mathrm{aD}$ (the lowest temperature after the die).

The gross and net $\mathrm{SME}$ ($\mathrm{kWh\ t^{-1}}$ or $\mathrm{J\ kg^{-1}}$) were determined according to the method described in \cite{Bastiaansen2023}, where the gross $\mathrm{SME}$ is calculated using the full-load current and the net $\mathrm{SME}$ is calculated by subtracting the idle-load current from the the full-load current. This enables us to differentiate between the total amount of energy dissipated (idle-load plus pellet extrusion) by the press and the amount of energy dissipated in the pellet extrusion process only. Only $\mathrm{SME_{net}}$ is used for the calculations in this manuscript.

\subsection{Physical pellet quality}
Pellet durability was determined using a Holmen NHP100 (Tekpro, USA) tester with a 3-mm screen. Approximately 100 grams of pellets underwent testing in the rotating drum mechanism, simulating mechanical stress \cite{Thomas1996}. The Pellet Durability Index (PDI) was calculated based on the fractional weight loss after tumbling $\mathrm{weight_{after}}/{\mathrm{weight_{before}}}$, representing the fraction of intact pellets after testing. A pellet durability close to $1$, indicates a high resistivity of pellets against abrasion during handling and transport. 

\subsection{Determining die residence time}
The total volume of the holes within the ring-die ($V_{\mathrm{die}}$) is calculated by summing ($n\times$) the volume of a single die hole ($V_{\mathrm{die\ hole}}=\frac{1}{4}D^2 L$). Subsequently, the characteristic residence time of the pellets within the ring-die ($t_{\mathrm{die}}$)  is estimated numerically by the ratio of the total ring-die hole volume ($V_{\mathrm{die}}$) and the volumetric flow rate of the ingredients through the ring-die ($\dot U$). The latter is determined by the mass flow rate ($Q$, the production rate) and the true density of a pellet inside a die hole ($\rho_{\mathrm{pellet}}$). 
We assume the true pellet density to be constant ($1200\ \mathrm{kg\ m^{-3}}$) for all pellets produced during this study. We validate this assumption based on the x-ray tomography density analysis presented by Benders \textit{et al} \cite{Benders2024}.
Therefore the characteristic residence time inside the ring-die is directly determined from $n$, $D$, $L/D$ and $Q$, where $Q$ is set by experimental tuning of the production rate and $n$, $D$ and $L/D$ are properties of the selected ring-die during the trial.

\counterwithin{figure}{section}
\section{Appendix}

\subsection{Controlling the moisture transfer through conditioning temperature}\label{app:sec:conditioning}
Inside the conditioner, significant heat and a small amount of moisture are transferred to the cooler particle mixture through steam condensation, raising both its temperature and moisture content. The increase in moisture content can be approximated based on the energy balance between the heat required to raise the temperature of the particle mixture ($m_{\mathrm{p}}\times c_p\times \Delta T_\mathrm{C}$) and the heat released by the condensation of steam ($m_{\mathrm{H_2O}}\times H_{\mathrm{vap}}$), as shown in Eq. \ref{eq:app_1}:

\begin{equation}\label{eq:app_1}
\Delta\mathrm{mc} = \frac{c_p\ \Delta T_{\mathrm{C}}}{H_{\mathrm{vap}}}
\end{equation}

In this equation, $\Delta\mathrm{mc}$ (the ratio between $m_\mathrm{H_2O}$ and $m_\mathrm{p}$) represents the change in moisture content due to steam condensation ($\mathrm{kg_{H_2O}\ kg_{p}^{-1}}$). The term $c_p$ is the specific heat capacity of the particle mixture ($\mathrm{J\ kg_{p}^{-1} K^{-1}}$), $\Delta T_{\mathrm{C}}$ is the temperature difference between the particle mixture entering ($T_{\mathrm{bC}}$) and leaving ($T_{\mathrm{aC}}$) the steam conditioner ($\mathrm{K}$), and $H_{\mathrm{vap}}$ is the latent heat of vaporization of water ($\mathrm{J\ kg_{H_2O}^{-1}}$).

The assumptions underlying Equation \ref{eq:app_1} are: (1) the contribution of latent heat exchange is much larger than that of sensible heat; (2) $c_p$ remains constant; (3) all particles reach a uniform temperature ($T_{\mathrm{aC}}$) during conditioning, such that there is no particle size effect; and (4) all heat is transferred from the steam into the particle mixture without losses to the environment. We discuss these assumptions and their effect on the experimental observations below.

\begin{figure}[!th]
\centering
\includegraphics[width=0.55\textwidth]{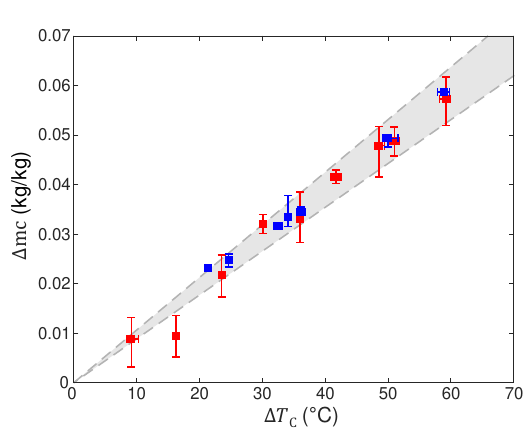}
\caption{\textbf{During steam conditioning, moisture and temperature changes are related through the energy balance between the energy required to heat the particle mixture and the energy released through steam condensation, where the slope is expressed as the ratio between specific heat capacity and the latent heat of vaporization.} Particle mixtures were conditioned using steam at a constant production rate, whilst varying the steam injection rate to induce changes in $\Delta T_{\mathrm{C}}$. The data represents two independent trials using the same composition, but from two different batches. In red: calculated data based on steam flow measurements ($Q_s/Q_p$). In blue: experimental moisture content based on $24\mathrm{h}$ oven drying at $105$ ${}^\circ \mathrm{C}$. The gray area visually represents the estimated moisture transfer according to Equation \ref{eq:app_1} by using $H_{\mathrm{vap}}=2.26\ \mathrm{MJ\ kg^{-1}}$, for values of $c_p$ between $2$-$2.4$ $\mathrm{kJ\ kg_{p}^{-1} K^{-1}}$.}\label{fig1:app}
\end{figure}

The change in moisture content during conditioning ($\Delta \mathrm{mc}$) can be determined through oven drying experiments using samples obtained before and after steam conditioning ($\Delta \mathrm{mc}= \mathrm{mc_{aC}}-\mathrm{mc_{bC}}$, see Fig. \ref{fig1:main}). Alternatively, monitoring the steam flow rate ($Q_s$) relative to the particle mass flow rate ($Q_p$) provides another way to estimate of moisture transfer during conditioning ($Q_s/Q_p$). As shown in Fig. \ref{fig1:app} and from Equation \ref{eq:app_1}, we observe a monotonic increase in moisture content with temperature change during conditioning, where the slope is set by $c_p/ H_\mathrm{vap}$. This consistency, regardless of the measurement method, suggests that our system injects high-quality steam, defined as the percentage of steam in the vapor phase \cite{Gilpin2002}. In this case, the latent heat of vaporization ($\approx 2.26\ \mathrm{MJ\ kg^{-1}}$) has a much larger contribution to the heating process than the sensible heat released by warm water per degree of cooling ($\approx 4.2\ \mathrm{kJ\ kg^{-1} K^{-1}}$), validating assumption (1).

The monotonic increase in $\Delta \mathrm{mc}$ as a function of $\Delta T_\mathrm{C}$ shows that moisture transfer is primarily driven by the temperature change during conditioning. Since the enthalpy of steam is constant, the specific heat capacity ($c_p$) of the particle mixture becomes the key ingredient-specific factor controlling moisture transfer. Thus, for a given $\Delta T_{\mathrm{C}}$, the amount of moisture transferred ($\Delta\mathrm{mc}$) can vary across mixtures due to differences in $c_p$, as suggested by Kulig \cite{Kulig2007}. The specific heat capacity itself can change during the process as ingredients undergo phase transitions, for example gelatinization or melting of starch \cite{Hwang1999}. The conditioning process is a process which takes place in $10$-$20$ seconds, during which very limited gelatinization of starch is observed \cite{Skoch1981,Stevens1987}. Consequently, we expect no major change in $c_p$ during the conditioning process according to assumption (2).

Assumptions (3) and (4) could not be validated experimentally. Nevertheless, we expect that heat transfer is not limited by the short processing times due to the rapid diffusion of heat into the ingredient mixture, relative to the typical particle diameter ($<2\ \mathrm{mm}$) \cite{Bouvier2014,Thomas2020}. If heat transfer were limiting the condensation process, we would expect a non-monotonic increase of $\Delta T_\mathrm{C}$ relative to the steam flow rate $Q_s$ at a constant production rate, as per the experimental conditions specified in Fig. \ref{fig2:main}. By increasing the injection rate, if the particles could not absorb the energy at the given rate, any excess steam would be lost to the environment. This would result in an increased steam injection rate relative to the inflow of particles $Q_s/Q$ (=$\Delta \mathrm{mc}$), and in $\Delta T_\mathrm{C}$. However, Fig. \ref{fig1:app} clearly shows a monotonic increase in moisture content with respect to the temperature change during conditioning. Furthermore, assuming $100\%$ thermal efficiency without losses to the environment (4) does not affect the shape of the results in Fig. \ref{fig1:app}, but it may contribute to an overestimation of the $c_p$ derived from the slope ($c_p\approx2.2\ \mathrm{kJ\ kg^{-1}K^{-1}}$). Based on literature, $c_p$ for our ingredient mixture is expected between $1.6$ and $1.8\ \mathrm{kJ\ kg^{-1} K^{-1}}$ (see \cite{Bovo2022,Ince2008,Otten1980,Kulig2007}), suggesting that we have overestimated thermal efficiency with assumption (4). Despite this overestimation, Eq.\ref{eq:app_1} still accurately represents the driving force for moisture transfer during steam conditioning.

In conclusion, the results highlight that the driving force during conditioning is the rapid condensation of steam on cold particles to reach thermal equilibrium. Unlike slower processes like vapor sorption, condensation allows feed producers to increase the mixture's temperature simultaneously with its moisture content in a matter of seconds.

\subsection{3D render of a pelleting die}
\begin{figure}[!th]
\centering
\includegraphics[width=0.55\textwidth]{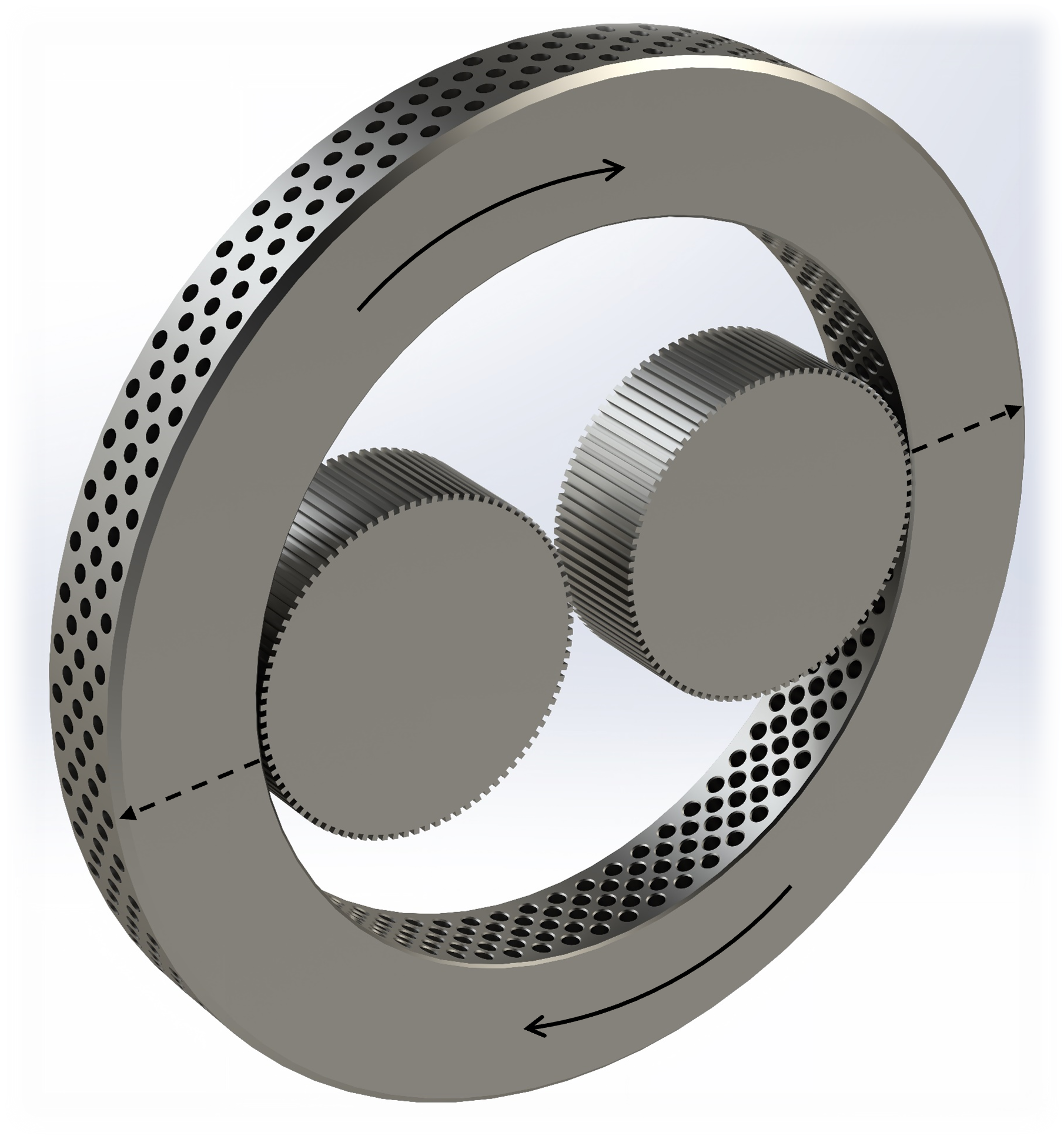}
\caption{\textbf{3D render of a pelleting die, equipped with two rollers with riffles.} The solid line arrows indicate the direction of the die's rotation, while the dashed arrows highlight the direction in which ingredients are transported through the die holes during pellet extrusion. Ingredients are caught in the gap between the roller and the die due to the die’s rotation. The roller’s riffles aid in gripping and guiding the ingredients into the die holes, where the material is consolidated and subsequently extruded to form durable pellets. }\label{fig2:app}
\end{figure}

\clearpage



\bibliographystyle{elsarticle-num.bst}
\bibliography{bibliography}






\end{document}